\documentclass[conference]{IEEEtran}
\IEEEoverridecommandlockouts

\usepackage{cite}
\usepackage{amsmath,amssymb,amsfonts}
\usepackage{algorithmic}
\usepackage{graphicx}
\usepackage{textcomp}
\usepackage{xcolor}

\usepackage[a4paper,width=150mm,top=25mm,bottom=25mm]{geometry}

\usepackage[nottoc]{tocbibind} % The tocbibind package can be used to add the ToC and/or bibliography and/or the index etc., to the Table of Contents listing

\usepackage[normalem]{ulem} % The ulem package provides various types of underlining that can stretch between words and be broken across lines.

\usepackage{multirow}

\usepackage{changepage}
\usepackage[export]{adjustbox}
\usepackage{rotating}

\usepackage{lscape}

\usepackage{amssymb}% http://ctan.org/pkg/amssymb
\usepackage{pifont}% http://ctan.org/pkg/pifont
\newcommand{\cmark}{\ding{51}}%

\usepackage{hyperref} % Provides LaTeX the ability to create hyperlinks within the document.
\hypersetup{
    colorlinks=true,
    linkcolor=black,
    filecolor=magenta,      
    urlcolor=black,
    citecolor=black,
}

\usepackage{nomencl} % The nomenclature package can be used to generate and format a nomenclature using MakeIndex.
\renewcommand{\nompreamble}{The next list describes several symbols \& abbreviation that will be later used within the body of the document}
\makenomenclature

\AtBeginDocument{%
  \providecommand\BibTeX{{%
    \normalfont B\kern-0.5em{\scshape i\kern-0.25em b}\kern-0.8em\TeX}}}

\title{An investigation of the Online Payment and Banking System Apps in Bangladesh\\
\author{\IEEEauthorblockN{1\textsuperscript{st} SHAHRIAR HASAN MICKEY}
\IEEEauthorblockA{\textit{(School of Data Science)} \\
\textit{Brac University}\\
shahriar.hasan.mickey@g.bracu.ac.bd}
\and
\IEEEauthorblockN{2\textsuperscript{nd} MUHAMMAD NUR YANHAONA}
\IEEEauthorblockA{\textit{(School of Data Science)} \\
\textit{Brac University}}
nur.yanhaona@bracu.ac.bd}
}
\begin{document}
\maketitle

\begin{abstract}

Presently, Bangladesh is expending substantial efforts to digitize its national infrastructure, with a significant emphasis on achieving this goal through mobile applications that facilitate online payments and banking system advancements. Despite the lack of knowledge about the security level of these systems, they are currently in frequent use without much consideration. To observe whether they follow the minimum global set standards, we choose to conduct static and dynamic analysis of the applications using available open-source analyzers and open-source tools. This allows us to attempt to extract sensitive information, if possible, and determine whether the applications adhere to the standards of MASVS set by OWASP. We show how we analyzed 17 .apks and a SDK using open source scanner and discover security flaws to the applications, such as weaknesses related to data storage, vulnerable cryptographic elements, insecure network communications, and unsafe utilization of WebViews, detected by the scanner. These outputs demonstrate the need for extensive manual analysis of the application through source code review and dynamic analysis. We further implement reverse engineering and dynamic approach to verify the outputs and expose some applications do not comply with the standard method of network communication. Moreover, we attempt to verify the rest of the potential vulnerabilities in the next phase of our ongoing investigation.

\vspace{1cm}
\textbf{Keywords:} MASVS, OWASP, Digitization Mobile App Security, Reverse Engineering, Payment Application

\end{abstract}

%%%%%%%%%%%%%%%%%%%%%%%%%%%%%%%%%%%%%%%%%%%%%%%%%%%%%%%%%%%%%%%%%%%%%%%%%%%%%%%%
\section{INTRODUCTION}

On 4th of February, 2016 \cite{bBHeist}, a number of money withdrawal requests, summing to 1 billion USD, were made from the Central Bank of Bangladesh, better known as the Bangladesh Bank, to its New York based Federal Reserve Bank. This turned out to be the biggest online bank heist attempt in history. Though majority of the requests were blocked due to suspiciousness in the requested messages, 81 mission USD nevere returned back to the country, which is still considered to be the largest amount till date. This one incident is enough to show the importance as well as the opportunity for investigation in the information and cyber security sector for business growth.\\

While IT adaptation is increasing at a rapid pace in Bangladesh in the financial and business sector after the Bangladesh Bank’s recommendation, most of the services are provided and consumed via smartphones through their applications, such as Nagad, bKash, Ibbl, Astha, Upay and many more. This is to avoid the everyday hassle of queuing in a line, minimizing the waiting time, and get the necessary services done - such as money transaction within the country, remittance transfer from abroad, utility bill payment, online purchase payment, Real Time Gross Settlement (RTGS) payment and other electronic fund transfers (e.g. BFTN/EFT) - at the blink of an eye while keeping the trace of such transactions digitally. The online mobile banking applications now provide the majority of the services 24/7 from anywhere in Bangladesh. This not only saves time for both the user and the service provider but also gives a new way of earning revenue \cite{impactstudy}.\\

The advancement in the implementation of the technology in the daily life chores component of local people in Bangladesh raises one of the most important but underestimated factors of how secure these applications are. There is no doubt that with the advancement of technology comes opportunities for growth \cite{parvez2021} but with that also comes negative aspects; e.g. cyber-threats. Indeed Bangladesh is prone to cyber attacks due to a developing country advancing towards digitization. Smartphone applications contribute to a large portion of the IT based infrastructure of Bangladesh since every now and then some form of mobile phone application interaction is taking place to carry out the daily chores of the local people. As of 2020, there were 170 million mobile phone connections serving 90 unique mobile subscribers among which 102 million mobile internet connections serving 47.1 million mobile internet subscribers \cite{gsma}.\\

\begin{figure}[htbp]
\centering
\includegraphics[width=0.5\textwidth]{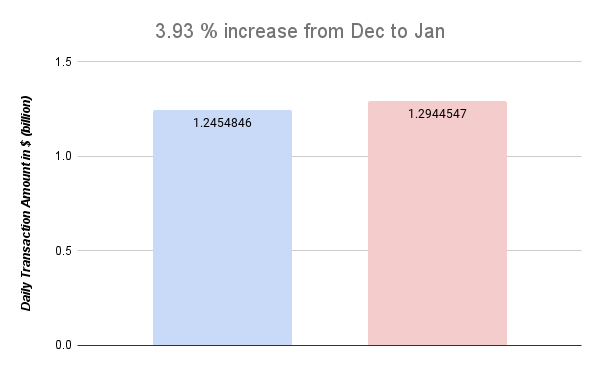}
\caption{Percentage increase in daily transaction amount from Dec to Jan}
\label{fig:Percentage increase in daily transaction amount from Dec to Jan}
\end{figure}

\begin{figure}[htbp]
\centering
\includegraphics[width=0.5\textwidth]{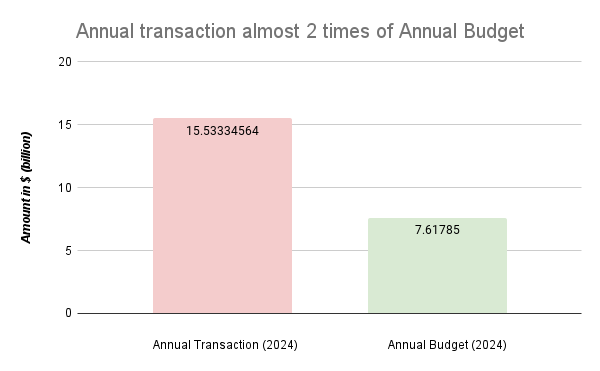}
\caption{Comparison between Annual transaction and Annual Budget of 2024}
\label{fig:Comparison between Annual transaction and Annual Budget of 2024}
\end{figure}

Point to be noted that the gap between the smartphone and the internet is routed by different types of mobile phone applications, most of which are service providing applications such as mobile banking, e-commerce and other financial services. These sudden and drastic changes are pushing Bangladesh to a new era of fin-tech where every type of task is somewhat dependent on the IT system. Just to give an idea of how big the Bangladeshi market of technology is, the Norwegian company, Telenor, receives 14.6 billion NOK (TK. 146 billion) just from Bangladesh alone. As of January 2024 \cite{bBank}, the total transaction rose to TK. 1,29,445.47 billion from TK. 1,24,548.46 billion in December 2023 which is a 3.93\% increase (See figure {\ref{fig:Percentage increase in daily transaction amount from Dec to Jan}}). But to understand how big this amount is, let us compare it with the budget of Bangladesh of the Fiscal year 2024 which is TK. 7,61,785 billion. This means the monthly transaction amount is almost one-fourth (1/6) of the Bangladesh budget, which becomes almost 2 times if calculated for annual transactions. The statistics could be visualized in the figure : {\ref{fig:Comparison between Annual transaction and Annual Budget of 2024}} . \\

Hence, our primary focus is on mobile application security, specifically the android system due to it being an open sourced environment. Not much work is done in the smartphone security sector in Bangladesh, specifically for Mobile Financial Services (MFS) system security. So far general review study on the online banking system and the security issues residing with it are discussed \cite{internetBanking, Akter2021}. However, neither any detailed analysis nor initiatives related to such concern has taken place. Thus, our research is on android application security related to this area, where we try to focus on the level of security issues, the possible vulnerabilities related to it, how the vulnerabilities could be exploited - if exist, detailed analysis of the android system and whether the service operating vendors in Bangladesh meet the global MFS application specification standards.\\

Lastly, let us talk about one type of UI based vulnerability research that takes control of the complete UI loop. A very specific type of attack like Clickjacking, which is the equivalent of the web vulnerability Tapjacking. In the past few years, a good amount of research work has been done to mitigate this vulnerability. Though this vulnerability is not completely uprooted since it is a feature of the android device itself which is used for many system level works by the android device. It is this feature of the android device that provides the permission SYSTEM\_ALERT\_WINDOW and BIND\_ACCESSIBILITY\_SERVICE that could be abused and exploited. If the malicious application is being downloaded from playstore then the user is not notified of the permission since the permission is granted by default. Hence draw on top is granted by default. This propagates to the possibility of advanced Clickjacking through “draw on top” and “accessibility service” permission. The Cloak and Dagger \cite{fratantonio17:cloakdagger} research work has demonstrated the issue with exceptional expertise and point of concern. Till date the vulnerabilities are exploitable and the study also shows how practical the attacks are without the user even understanding what has actually happened.

% \newpage
\section{\textbf{Problem Statement}}

In Bangladesh, the widespread use of mobile applications for online payments and banking systems raises concerns about their security. Given the lack of knowledge about their security levels and adherence to the global security standards, this research conducts static and dynamic analysis of 17 android applications and a SDK using the available open-source tools. The findings reveal security flaws, including data storage weaknesses, vulnerable cryptographic elements, insecure network communications, and unsafe use of WebViews, highlighting the need for comprehensive manual analysis and the importance of ensuring standard–compliant network communication in these applications.\\

The rest of the chapters of the report are organized as follows. We give an overview of the related works that have been done in the field both locally and globally in Section \textcolor{red}{III}. Following that we further discuss in detail  regarding our designed methodology of the analysis in Section \textcolor{red}{IV} \& \textcolor{red}{V} and describe the experimenting applications and SDKs in Section \textcolor{red}{VI}, along with the findings in Chapter \textcolor{red}{VII}, \textcolor{red}{VIII} \& \textcolor{red}{IX}. Finally, in Section \textcolor{red}{X} \& \textcolor{red}{XI} we talk about the scope for future work and conclude our work.
\\

\section{\textbf{Related Work}}
Many malicious android applications take the advantage of permission groups and try to misuse the power and privilege to exploit the vulnerabilities present in certain API versions of the device. Bartel et al. \cite{bartel} uses a comparison approach to see whether the requested permission to the users are actually required within the application itself. Similarly, VetDroid \cite{vetdroid} was a research output which was aimed to discover and examine undesirable behaviors within android applications, by analyzing the process of accessing sensitive system resources after taking sensitive permissions, and how those resources are further used by the application. This easily allowed security analysis to be made over the application inorder to vet the internal sensitive behaviors of the application in custody. \\

Following towards more recent works, to protect against permission-abuse attacks in android devices, inspired from above two works, DyPolDroid \cite{dypoldroid} uses a dynamic and semi-automated security framework which is based upon android enterprise, to give users and organizations the privilege to shield against malicious leverage of permission for malicious purposes, even if the user does not have any expertise in such field. DroidCap \cite{droidcap}, however, uses an OS-level approach to support so-called capability-based permissions in android, that provides further separation of privileges within an application by modifying the Android Zygote and IPC. But, to achieve a more specific approach in defeating malicious applications, OS-level modification is required in this case. \\

Moreover, research work such as BorderPatrol \cite{BorderPatrol} and Reaper \cite{reaper} make use of some unique techniques to fight against the malicious application. While the former attempts to protect the device by generating a customized Mobile Device Manager which leverages fine-grained contextual information, the later provides a real-time analysis of the android application, augmenting and complementing the android permission system, hence potentially countering the ongoing attack. However, both of them use the Xposed Module Repository \cite{xposed}, they require root level access, making it more complicated in terms of user friendliness. GDdroid \cite{gdroid} on the other hand, makes use of neural networks to map applications and APIs into a graph, translating app classification into node classification. Inorder to check API behavior patterns, they model an embedding-based method. \\

Research work such as FlowDroid \cite{flowdroid}, which work on the taint analysis of android applications, and Amandroid \cite{amandroid}, which is based on component flow in context-sensitive way work, are some of the novel work which are pivotal point for the generation of many other prominent research works. Besides taint and component analysis, intent based analysis is another approach to tackle the security issues in android applications. One such work is Epic \cite{epicc}, that computes android Intent call parameters using the IDE framework, similar to that used in FlowDroid. It does so by modeling the intent data structure explicitly in the flow functions. \\

In the state-of-art of android application static analysis, IccTA \cite{iccta} and DroidSafe \cite{droidsafe} made some major progress. While IccTa \cite{iccta} track data flows through regular Intent calls and returns by leveraging the work from FlowDroid, DroidSafe tracks both Intent and RPC calls. TaintDroid \cite{taintdroid} is however a dynamic runtime taint-tracking system that attempts to look for potential misuse of a user's private information. The dynamic analyses present in the system are all based on evasion attacks. \\

However, such works are done based on mainly plain APKs that do not use any obfuscation techniques. However, the real problem arises when there is the presence of code obfuscation at the source code level which makes it useless to use such tools. Also not much work has been done in this part of the android application. There is also some UI level research work which provides deep study analysis on topics such as tapjacking \cite{tapjacking}, \cite{fratantonio17:cloakdagger} and how to mitigate such security issues.\\

\subsection{\textbf{Tools for security analysis}}
\label{tools}

There are numerous tools available for enhancing the process of static analysis on APK and AAR files which are commonly used in android application development. Some of the popular names for static analysis tool chains are:

\begin{enumerate}
    
    \item AndroGuard \cite{androguard} : This open-source tool is specially designed for analyzing android applications. It offers a wide range of features, including APK information extraction, bytecode analysis, vulnerability detection, malware analysis, as well as generation of call graph.
    
    \item JADX \cite{jadx} : This tool serves as a decompiler and static analyzer for APK files. It allows developers to decompile APK files into readable Java source code, making it easier to analyze the application’s behavior, logic, and potential security issues.
    
    \item MobSF (Mobile Security Framework) \cite{mobsf} : MobSF is an open-source framework designed for mobile application security testing. It supports static analysis for both APK and AAR files. MobSF offers features such as vulnerability scanning, privacy testing, and code reviewing. This is based on a Javascript platform which uses a local server to run the scanner and review the report of the scanned output. 
    
    \item AARDroid \cite{aardroid-acsac22} : AARDroid is a static analysis tool which is a result of a research paper designed to identify security vulnerabilities in android applications. It focuses on detecting weaknesses, such as the storage of unencrypted credit card information in files, usage of insecure cryptographic algorithms, insecure methods for inputting credit card information, and insecure utilization of WebViews, in the application.
    
    \item APLTools \cite{apktool} : It is an APK decompiler and recompiler which helps to extract the source code by decompiling it as SMALI code, make changes to the source code and recompile it to run on the device with the modified version.
    
    \item Radare2 \cite{radare2} : “radare2” (often stylized as ‘r2’) is a powerful open-source framework for reverse engineering and analyzing software binaries. It is a command-line tool that provides a wide range of features for disassembling, debugging, analyzing, and manipulating the binary files, including  executables, libraries, firmware, and many more.
    
    \item Ghidra \cite{ghidra} : A software reverse engineering (SRE) framework developed by the National Security Agency (NSA). Ghidra is used for analyzing and exploring compiled code to understand its functionality, vulnerabilities, and behavior.
    
    \item QARK (Quick Android Review Kit) \cite{qark} : It is an open-source security auditing tool developed by LinkedIn. It focuses on identifying security vulnerabilities, insecure coding practices, and potential information leaks in android applications through static analysis.
    
    \item AndroBugs \cite{androbugs} : Androbugs is an open-source Python-based statistical analysis tool for android applications. It scans APK files, examines the manifest, and analyzes the source code to identify potential security vulnerabilities, privacy issues, and insecure coding practices.
    
    \item SonarQube \cite{sonarqube} : SonarQube is a widely used code quality and security analysis platform that supports the analysis of android projects. It can be integrated into the development pipeline and used to detect security vulnerabilities, bugs, and code smells within the APK and AAR files.
    
    \item Fortify Static Code Analyzer \cite{fortify} : Fortify is a commercial static code analyzer that provides comprehensive security analysis for various programming languages, including Java used in Android development. It offers advanced capabilities for analyzing APK and AAR files to identify security vulnerabilities, coding flaws, and potential weaknesses.

\end{enumerate}
	Besides the static analysis frameworks, there do exist tools that help in carrying out the dynamic analysis of the application. Some of them are as follows:

\begin{enumerate}
    
    \item Frida \cite{frida} : It is a dynamic analysis toolkit which makes tasks easy for researchers, developers as well as reverse engineers to do dynamic analysis. It is actually a scriptable tool that helps to hook on to different classes and components of the android application at runtime so that application states could be traced to perform different tasks.
    
    \item Burp suite \cite{burpsuit} : Burp suite is a well known intercepting tool in the network industry which has its equal share of usage in the web application as well as mobile application sector. Along with many other features, it primarily helps to intercept the network communication, working as a proxy between the server and the client, providing a plethora of features and services to do dynamic network communication analysis.

\end{enumerate}

\subsection{\textbf{Global Security Standards}}

	Application architectural global standards exist to ensure consistency, compatibility, and uniformity across various fields like technology, manufacturing, safety, and quality. They enable different products and systems to work seamlessly, reduce costs by streamlining processes, maintain quality through guidelines, enhance safety, foster  innovation, and facilitate global trade. Among other standards, two of the most prominent and widely used industry standards, for their specialization and coverage, are MASVS \cite{masvs} and PCI-DSS \cite{pcissc}.

\subsubsection{\textbf{MASVS}}
\label{owasp-masvs}
The OWASP MASVS (Mobile Application Security Verification Standard) \cite{masvs} is an extensive framework created by OWASP to assist organizations in establishing and upholding the security in their mobile applications. It offers a range of guidelines and suggestions for secure mobile application development, testing and deployment. The objective of MASVS is to aid organizations in constructing mobile applications with strong security measures and safeguarding them against prevalent vulnerabilities encountered in mobile applications. OWASP, a nonprofit organization, created MASVS to address the unique security challenges presented by a community of security experts, developers and researchers with the objective of promoting best practices and improving the security posture of mobile applications.\\

The primary objective of MASVS is to help organizations build and maintain secure mobile applications. It offers a comprehensive set of requirements and recommendations, spanning various security aspects such as authentication, data storage, network communication, cryptography and secure coding practices. MASVS assists organizations in identifying and addressing potential security weaknesses throughout the mobile application development lifecycle.

\subsubsection{\textbf{PCI-DSS}}

The PCI-DSS (Payment Card Industry Data Security Standard) \cite{pcissc} was developed by leading credit card companies like Visa, MasterCard, American Express, Discover, and JCB. Its purpose is to safeguard cardholder’s data and ensure secure transactions within the payment card industry. It traces back to the creation of PCI Security Standards Council in 2006 where this collaborative effort among card brands aimed to bolster cardholder‘s data security and combat the increasing threat to data breaches. At its core, PCI-DSS aims to protect the sensitive information of the cardholder. It outlines comprehensive security requirements that must be followed by the organizations involved in payment and transactions. These requirements encompass areas such as network security, encryption, access control, vulnerability management, and information security policies.\\

Now talking about the implementation of PCI-DSS, it involves several steps. Organizations must assess their payment card processing environment to determine the applicable requirements based on transaction volume and specific card brands. Compliance to validation methods include self-assessment questionnaires (QSAs), quarterly network scans by Approved Scanning Vendor (ASVs), and periodic penetration testing. Failure to comply with PCI-DSS can result in significant consequences. Organizations may face fines, penalties, reputational harm, heightened risk of data breaches, and potential suspension of card processing privileges. Thus, ongoing compliance and robust security measures are crucial to protect cardholder’s data and mitigate risks associated with it. \\

Smartphone and its application vulnerabilities encompass a range of weaknesses in mobile applications, operating systems \cite{adascalitei, alzadjali}, and user behaviors that can be exploited by attackers \cite{amarante}. These vulnerabilities can lead to unauthorized access, data breaches, and compromise user privacy and security. Thus it is crucial for mobile developers, users, and the organizations to stay vigilant, apply security best practices, and keep their devices and applications up to date to mitigate the risks associated with these vulnerabilities. These vulnerabilities could be outlined as follows : operating system vulnerabilities, App Store Security, Data Storage, Network Communications, Mobile Malware, Device Configuration and User Behaviors. All of these categories would take years of research to exploit as each of them in itself has their own sub fields taking years of experience to explore. \\

In order to make our task simple, we focused on the TOP 10 most common vulnerabilities according to OWASP as the base \cite{top10}. If the vulnerabilities mentioned in section 3.1 exist in an application then there is the possibility that attacks can be made, for example, through reverse engineering \cite{reverseengineering}, side channel attack \cite{side_channel}, behavioral manipulation \cite{dypoldroid}, along with many other techniques.

\section{\textbf{Research Methodology}}
\label{researchMethodology}

Among the tools mentioned above in Section \textcolor{red}{\ref{tools}}, there exist both open source as well as subscription based applications; the subscription based ones grants access to updates and features from the vendors at a cost of services charge. Regardless of the tools being open-sourced or commercial, these tools aid in static analysis of APK and AAR files, aiding developers in pinpointing and addressing the security concerns. However, our focus primarily centers towards the open-source options to make the work methodology easily accessible and tools readily available, ensuring reproducibility of the work. \\

Our primary concern is on the android application security where we try to focus on its possible vulnerabilities, how the vulnerabilities could get exploited, how the application could get affected by the exploitation of the vulnerabilities, and to what extent the vendors, operating in Bangladesh, meet the global application standards set by the MASVS. We try to give a detailed analysis of the investigation of the android system we perform which is mentioned in Section \ref{sec:investigation}. Inorder to get full leverage, we must first understand the types of possible vulnerability regions. We classify them by leveraging the approach used to classify in the research work AArdroid \cite{aardroid-acsac22}. They are as follows:

\begin{enumerate}
    \item\textbf{Unsafe Usage of WebView} pertain to security weaknesses found in the implementation and usage of WebView components within web browsers or mobile applications. WebView is an Android system component that enables the embedding of web content in applications.

    \item\textbf{Weak Cryptographic Operations} refer to weaknesses and flaws in the implementation, configuration, or usage of cryptographic algorithms and protocols. These vulnerabilities can jeopardize the security of sensitive data and cryptographic operations.
    
    \item\textbf{Security Vulnerabilities with User Interaction} are security weaknesses and flaws that can be exploited through user interactions or inputs, potentially compromising the application's security and leading to unauthorized access or actions.
    
    \item\textbf{Insecure Networking} pertain to security weaknesses and deficiencies that can be taken advantage of via network communications. These vulnerabilities have the potential to enable unauthorized access, data breaches, or the compromise of sensitive information.
\end{enumerate}

Presence of such types of vulnerabilities means severe consequences in cases they could get exploited. This could lead to data leakage, identity theft, privilege escalation to compromise systems, data and credential theft and many more. So in order to tackle such issues we have to work towards mitigating them. Hence to ensure mitigation we need to stick with the set standard for industry best practices. One such standard is MASVS by OWASP, as mentioned in Section {\ref{owasp-masvs}} in brief.  \textbf{Implementing MASVS} involves incorporating its guidelines which are thorough and very much contentful, but some of the contents do not apply for many of the applications. From the 6 types of standards mentioned there we work with 4 of them in our work, each consisting of its own subcategory. We are stating the categories with their subcategories as follows:\\

MASVS-STORAGE:
\begin{enumerate}
    \item The application securely stores crucial data 
    \item The application prevents leakage of important data
\end{enumerate}

MASVS-CRYPTO:
\begin{enumerate}
    \item The application should consider recent strong cryptographic methods and should comply with industry best practices
    \item The application does key management complying with industry best practices.
\end{enumerate}

MASVS-NETWORK:
\begin{enumerate}
    \item The application secures network traffic complying with present best methods
    \item The application performs identity pinning for every remote endpoints within the developer's supervision.
\end{enumerate}

MASVS-PLATFORM:
\begin{enumerate}
    \item IPC (inter process communication) mechanism is used securely by the application
    \item Web Views adopted securely by the application
    \item User Interface is adopted securely by the application
\end{enumerate}

% \newpage
\section{\textbf{Investigation Process:}}
\label{sec:investigation}

We initially perform static analysis of the application through source code and manifest file review. Inorder to do so we would require to decompile the APK file first since the application being dealt with are from google play store, hence we would not be able to get the source code from any other open source system. We will be requiring the following set of tools for our investigation :

\begin{enumerate}
    \item Apktool
    \item Jadx
    \item Android Studio
    \item MobSF
    \item AArdroid
    \item Google enabled API emulator / Android Device
    \item Keygen
    \item Jarsigner
    \item Align tool
    \item ADB (android debug bridge)
\end{enumerate}

Apktool \cite{apktool} and Jadx \cite{jadx} are two of the most important tool sets for us. Briefing about the tools are mentioned in Section {\ref{tools}}. We first attempt to decompile the APK using Apktool to get the structure of the zipped application file and get the source code and other static resources along with the manifest file. But since the decompiled source code is actually in the format of android / java byte code it would not be a lot of help if we try to do source code review on that byte code, as doing so would not grant us true progress any faster. Hence we are required to implement reverse engineering techniques to trace back and regenerate the equivalent Java code of that byte code. The byte code is actually like assembly language, also known as Smali code, for the android runtime system. Now working with the Java source code is much easier than the Smali code. To understand the structure and the working and operational principle of the Smali code we will need to understand the opcodes \cite{google_android_1, google_android_2} specific to the Smali code so that we can then write our own desired codes in the decompiled applications.\\

\begin{figure}[htbp]\centering
\includegraphics[width=0.5\textwidth, inner]{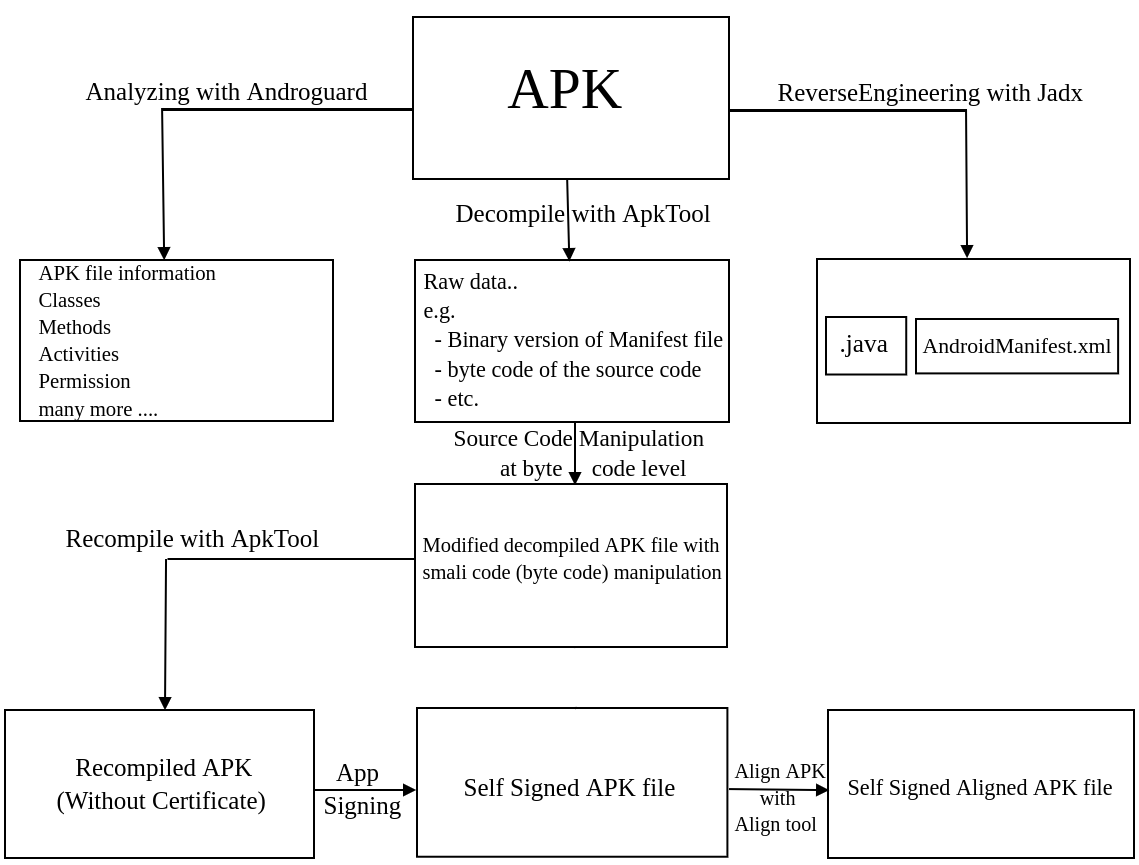}
\caption{Reverse Engineering FlowChart}
\label{fig:Reverse Engineering FlowChart}
\end{figure}

Besides working with the decompilation of files and source code analysis, we would also require Android Studio \cite{androidstudio}, which is the preferred IDE for android application development by google. The reason for using android studio is to ensure to put the correct code manipulation to source code to the target application. Since Smali itself is a very difficult language and the documentation of the Smali code is not that great, Android Studio is the next best alternative to work with and get the almost perfect Smali equivalent code. First we need to create a dummy project in Android Studio and write the required code we want to include in the target APK into the dummy project. Then the project is compiled to generate the APK file. Finally, Apktool is used to decompile the dummy APK. From the decompiled version we can now get to the Smali source code and get to the class file where we have our required statement. Using any code editor or Android Studio or even from the terminal we can open the file and look for the Smali version of our desired statement, copy it and paste it into the target application. Lastly we recompile the modified decompiled APK file using Apktool, install it into an emulator or real android device and run the application in the physical or virtual device. But before being installed, we need to first generate a public-private RSA key pair using keygen, sign the recompiled APK with Jarsigner and finally align it using the Align tool. The signing and aligning tool can be found in any linux terminal. The reverse engineering process is summarized in the flow chart in fig : {\ref{fig:Reverse Engineering FlowChart}} \\

\begin{figure}[htbp]\centering
\includegraphics[width=0.5\textwidth]{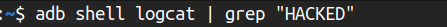}
\includegraphics[width=0.5\textwidth]{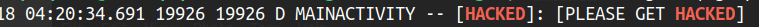}
\caption{Logcat Data}
\label{fig:logcat}
\end{figure}

In our case, we wanted to insert log statements into the target APK to check from where and when certain specific activities and classes get called. For ensuring it worked we inserted a log statement to the MainActivity of the target APK and the desired output at initiating the application is shown in Fig: \textcolor{red}{\ref{fig:logcat}}\\

\begin{figure}[htbp]
\centering
\includegraphics[width=0.5\textwidth]{core/logcat_cmd.png}
\includegraphics[width=0.5\textwidth]{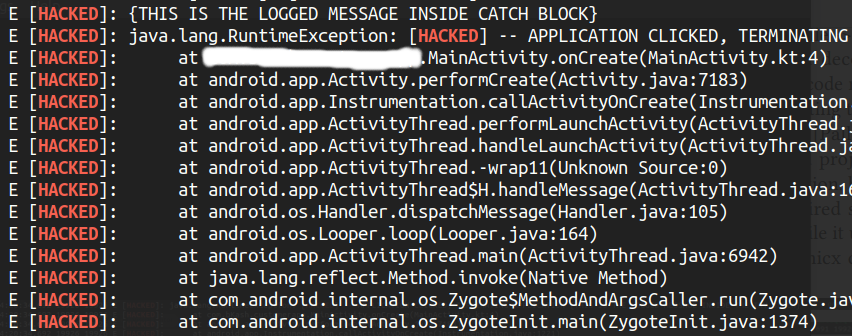}
\caption{Logcat Data}
\label{fig:logcat2}
\end{figure}

Since an android device has its own shell and the log statements are written to a log file inside the android system, we need to use logcat in the android device using adb shell. However, the log file does not only contain our log statement, rather it is the dumping ping for all the logs of the entire android device. So by running only the command to see the log of the device our terminal will be flooded with hundreds, if not thousands, of lines of log statement, from which it will be very difficult to look for our log statement as all the log statements are dumped at the same time. To get rid of this issue we can try to filter the output of the log file using the grep tool in the terminal. In our case we further want to insert a statement to throw an exception whose output is shown in Fig: Fig: \textcolor{red}{\ref{fig:logcat2}}. The reason for inserting exception statement is to keep track of all the classes and method which are being called at runtime of the application so that we can generate a call graph sequence and which will also help us later in our research to look into the state of the variables and parameters at specific time of the execution. \\

\begin{figure}[htbp]
\centering
\includegraphics[width=0.5\textwidth]{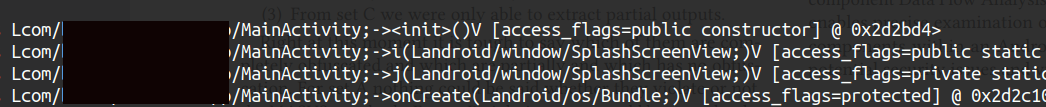}
\caption{Androguard analyzer}
\label{fig:androguard_analyzer}
\end{figure}

To collect and list down the classes present under a specific package, the methods present inside a class, or to show the state of the AndroidManifest file by grouping each category into separate groups will make the investigation work faster. Androguard \cite{androguard} is a handy tool that can do all of this detailing by organizing the information of the APK and show only those information which are demanded. For example we try to look into the methods of MainActivity class. The output is shown in Fig: {\ref{fig:androguard_analyzer}} \\

However, working with the source code of any production level application for doing manual static source code analysis is not very feasible since such an application contains millions of lines of codes. Hence it is much more helpful to use any sort of automated scanner that would parse the whole application and look for the common source to sink links to find regions of potential vulnerabilities. Talking about scanners, there are not many open source scanners that are designed to effortlessly scan the APK files. Among the very few, two of them are MobSF \cite{mobsf} and AArdroid \cite{aardroid-acsac22}. We will be using both of them. MobSF is good for the low hanging fruits while Aardroid is better for finding the presence of specific types of potential vulnerabilities we are focusing on. We run MobSF initially and the result of scanning is shown on a web browser at the specified url that it provides during the scanning at the terminal. There we get to see if any specific type of key is hard coded to the application source code or whether any database or other authentication related data is stored with encryption or not along with other results. \\

Now shifting our focus towards Aardroid, it is specifically made to work for AAR or android archive file which is actually an android library that could be used as dependency in any android project. One example of such a library could be a payment gateway library for banking and MSF applications. What Aardroid does is to make a dummy project and in that it keeps the target AAR file as a dependency. Then it attempts to compile the android project, generating an APK file. Finally it uses Apktool to decompile the APK. After decompilation it does its analysis on finding the presence of the vulnerabilities in the APK due to the integration of that AAR file as dependency using Androguard, Cryptoguard and a parser of itself. In our case we only have one AAR file in our target application set. That means the raw Aardroid is not going to perform its analysis in a prebuilt APK. So Aardroid itself is not of much help to us. But we can try to bypass the APK creation and the toolchain from the APK analysis directly. This is the contribution of our research work. The customization we make is on the next page:\\

% \newpage

\textbf{\textit{Customization in Aardroid: }} 
\begin{enumerate}
    \item If the target application is an AAR file then build the dummy APK with the necessary components required for the analysis
    \item Else if the file is APK then no need to generate the APK, rather decompile the APK and start the analysis directly
        \begin{enumerate}
            \item If the APK is not decompilable then simply end the process and move on to the next application in the target application dataset
            \item Else if decompilable then after decompilation of the APK:
                \begin{enumerate}
                    \item If decompilation is successful and generation of call graph using Amandroid (a.k.a. ArgusSAF) then execute Amandroid and Cryptoguard
                    \item Else if call graph cannot be generated using Amandroid then:
                        \begin{enumerate}
                            \item If Cryptoguard is runnable on the application then successfully generate output of partial analysis (This is able to only do the analysis of the Platform Interaction properties from PLAT2 to PLAT8 type of vulnerabilities as mentioned in the last 7 columns of table {\ref{tab:plat1_to_plat7_chart}})
                            \item Else unsuccessful attempt with both Amandroid and Cryptoguard
                        \end{enumerate}
                \end{enumerate}
        \end{enumerate}
\end{enumerate}

The customization could be visualized in the flow chart below {\ref{fig:customized aardroid}}: 

\begin{figure}[htbp]
\centering
\includegraphics[width=0.5\textwidth]{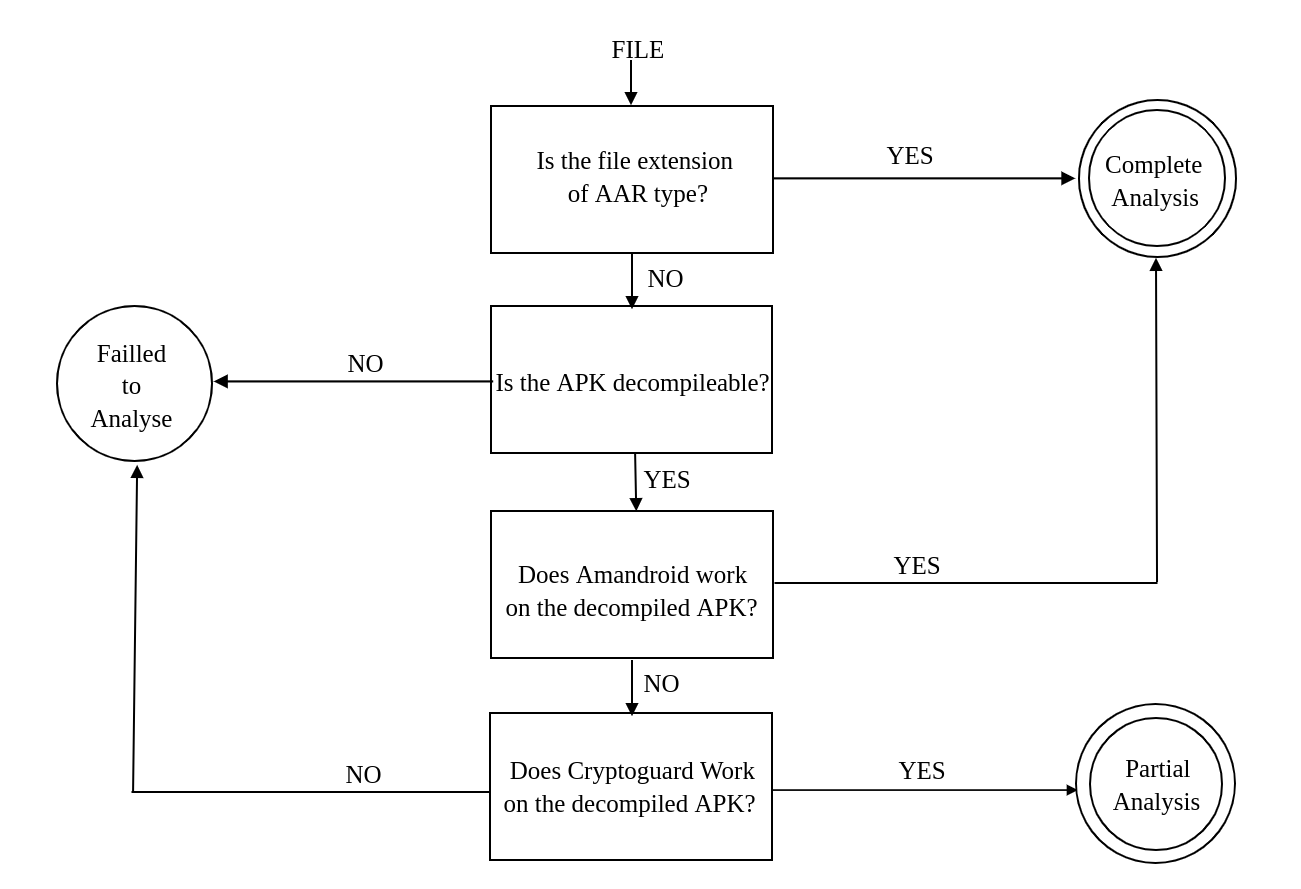}
\caption{Customized Aardroid}
\label{fig:customized aardroid}
\end{figure}

Lastly, emulators are preferred over physical android devices since emulators provide extra added benefits as well as it helps to make sure whether sensitive applications work on rooted devices or not. Moreover, keeping more than one android device is not expected as well as not feasible to carry out experimentation. We have the option to use emulators from the following images : 
\begin{enumerate}
    \item Google emulator
    \item Without Google API Services emulator
    \item With Google API Services emulator
\end{enumerate}

We try to use all of them, however, option (1) and option (3) does not allow root access. To gain root access individually on option (3) we use rootAVD \cite{rootAvd}. Hence root privilege can be achieved. However, we try to conduct our investigation without taking root access, keeping in mind that production level applications will not always run on rooted devices. Hence we considered keeping the developer mode to be enabled only. We discuss the challenges we have faced throughout the investigation in Section {\ref{challenges}}.

\section{Testing Application \& Architecture}

We chose the following set of applications as our testing targets:

\begin{table}[h]
    \centering
    \begin{tabular}{|c|l|c|}
    \hline
    No. & SDK / APK Names & Category \\
    \hline
    1   & bKash          & \multirow{3}{*}{MFS / Banking Wallet}\\
    2   & Nagad          &  \\
    3   & Rocket         & \\
    % 4   & Cellfin        & \\
    \hline
    4   & City DRF-SCF   &  \multirow{11}{*}{Mobile Banking App}\\
    5   & CityRemit      & \\
    6  & Southeast Bank Mobile App & \\
    7   & NexusPay       & \\
    8   & Islamic Wallet & \\
    9   & Astha by Brac Bank &  \\
    10  & IFIC Amar Bank & \\
    11  & ibbl-ismart by Islami Bank & \\
    12  & UniON Bank     & \\
    13  & ONE Bank App   & \\
    14  & UPay           & \\
    \hline
    15  & MY-GP App      & Mobile Recharge App \\
    \hline
    16  & DWASA          & \multirow{2}{*}{Utility Bill Payment} \\
    17  & Desco          & \\
    \hline
    18  & SSL Commez     & Payment SDK \\
    \hline
    \end{tabular}
    \caption{Target Applications set. (No. 18 is an Archive file).}
    \label{fig:experimenting_apk}
\end{table}

From 1 to 3 are the top three online MFS applications in Bangladesh. The following 11 applications are banking apps providing MFS along with other applications providing other services with moderate but comparatively lower user base than the first three. CityRemit, which is at no.5 is basically for sending remittance to Bangladesh from abroad. This is a product of City Bank. At 15 is myGP app which is for sim recharge and internet buying app by Grameen Phone, the Bangladeshi version of Telenor. 16 and 17 are governmental applications for water and electricity bill payment. At 18 is SSL Commerz’s AAR file which is a payment SDK. It is needed to be integrated into an application in order to provide the payment gateway to different e-commerce applications for making purchase payments. \\

Table {\ref{fig:Obfuscated vs Not Obfuscated applications}} distinguishes between the obfuscated and the unobfuscated applications. For the convenience of this research we are also considering those applications which have the XAPK format as obfuscated application.

\begin{table}[h]
    \centering
    \begin{tabular}{|c|l|c|l|c|}
    \hline
    No.  & Obfuscated & Not Obfuscated \\
    \hline
    1            & \cmark &\\
    \hline
    2         &  & \cmark\\
    \hline
    3         & \cmark &\\
    \hline
    % 4  & Cellfin          &  & \cmark\\
    % \hline
    4  &  & \cmark\\
    \hline
    5      & & \cmark\\
    \hline
    6   & & \cmark\\
    \hline
    7        & & \cmark\\
    \hline
    8  & & \cmark\\
    \hline
    9   &  & \cmark\\
    \hline
    10  & & \cmark\\
    \hline
    11   &  & \cmark \\
    \hline
    12     & \cmark &   \\
    \hline
    % 13  & ONE Bank App   & &\\
    13    & & \cmark\\
    \hline
    14         & \cmark &\\
    \hline
    15      &  & \cmark\\
    \hline
    16        &  & \cmark\\
    \hline
    17          &  & \cmark\\
    \hline
    
    % \hline
    % 18  & SSL Commez     &  &\\
    % \hline
    \end{tabular}
    \caption{ Obfuscated vs Not Obfuscated applications.}
    \label{fig:Obfuscated vs Not Obfuscated applications}
\end{table}

It can be observed from the table above that very few applications follow the strategy of source code obfuscation. This in itself is a security measure that, if not secure completely, makes it time consuming and difficult to understand and break into. But the concerning fact is that majority of the applications are not obfuscated which marks it as weak point of the application since the applications are now like a open book; so any malicious user / attacker, with sound knowledge in the application security, can try to review the source code by reverse engineering and try to exploit any potential vulnerabilities to extract data or compromise the system by using unfair means. \\

The sole reason for considering these applications for our testing set is that they are the most widely and regularly used applications in Bangladesh that deal with financial services. Deficiency in security and compromise to any of the systems mean a huge loss in the economic stability of Bangladesh. Hence, our motivation leads us to funnel through the huge set of applications and get the residue of the target application shown in table {\ref{fig:experimenting_apk}}.\\

The table {\ref{tab:Required permission by target applications}}  summarizes the permissions each of the applications ask for. There are also some custom permissions that majority of the applications ask for, which we have not disclosed here for security reasons since they are specific to the vendors. \\

\section{Findings}

\textit{\textbf{Details on scanned results of Aardroid:}}\\
After subjecting all 17 APKs and 1 SDK to the analysis process, we have obtained our scaned outcomes. Yet, it's worth noting that not all of the applications were decompilable using Aardroid. Additionally, the APKs are pre-existing and encompass a substantial amount of UI-oriented activities and classes, which we directly processed through the scanner; hence, the analysis procedure consumed a considerable amount of time. \\

The results could be divided into 3 sets:
\begin{enumerate}
    \item Set A : Applications that Aardroid is unable to decompile, hence was not able to scan and automate the analysis
    \item Set B : Those which are partially analyzable using Aardroid
    \item Set C : Complete scanning using Aardroid

\end{enumerate}

\vspace{12pt}

The output from Aardroid is split into 3 different tables (Table \textcolor{red}{\ref{tab:ds1_to_ds12_chart}}, \textcolor{red}{\ref{tab:crypto1_to_tls4_chart}} and \textcolor{red}{\ref{tab:plat1_to_plat7_chart}}) for convenience and better understandability. Table \textcolor{red}{\ref{tab:ds1_to_ds12_chart}} represents output related to \textbf{Data Storage and Privacy (DS)}, table \textcolor{red}{\ref{tab:crypto1_to_tls4_chart}} represents \textbf{Cryptography (CRYPTO)} and \textbf{Network Communication (TLS)} related properties while table \textcolor{red}{\ref{tab:plat1_to_plat7_chart}} shows \textbf{Platform Interaction (PLAT)} oriented properties. (\textbf{Those which are marked with "\textcolor{red}{V}" violates the property, while those with "N" does not show any violation. However, "N/A" represents that the property is not applicable to that application and "-" states that aardroid is unable to verify}).  \\
\\
\textbf{\textit{NOTE : }} For space limitation, we have dropped the data of Set A due to the reason that Aardroid could not scan the applications, hence, nothing could have been inferred about them. We also dropped the data of set B from DS1-DS12 in Table \textcolor{red}{2} and CRYPTO1-TLS4 in Table \textcolor{red}{3} for the same reason. Set A consists of applications no. 4, 6, 9, 13 and 14 while Set B consists of application no.s 2, 12, 16, and 17.\\

\begin{table*}[h]
    \begin{adjustwidth}{}{}
    \begin{adjustbox}{width=15cm}
    \begin{tabular}{|c|c|c|c|c|c|c|c|c|c|c|c|c|}
      \hline
        \textbf{App No.} & 
        \textbf{DS1} &
        \textbf{DS2} &
        \textbf{DS3} &
        \textbf{DS4} &
        \textbf{DS5} &
        \textbf{DS6} &
        \textbf{DS7} &
        \textbf{DS8} &
        \textbf{DS9} &
        \textbf{DS10} &
        \textbf{DS11} &
        \textbf{DS12} \\
      \hline
      1 & N & N & N & N & N/A & N & N/A & \textcolor{red}{V} & N/A & N/A & \textcolor{red}{V}(L) & \textcolor{red}{V}(L) \\
      \hline
      % 2 & - & - & - & - & - & - & - & - & - & - & - & - \\
      % \hline
      % 3 & - & - & - & - & - & - & - & - & - & - & - & - \\
      % \hline
      % 4 & - & - & - & - & - & - & - & - & - & - & - & - \\
      % \hline
      5 & N & N & N & N & N/A & N & N/A & \textcolor{red}{V} & N/A & N/A & \textcolor{red}{V}(L) & \textcolor{red}{V}(L) \\
      \hline
      % 6 & - & - & - & - & - & - & - & - & - & - & - & - \\
      % \hline
      % 7 & - & - & - & - & - & - & - & - & - & - & - & - \\
      % \hline
      8 & N & N & N & N & N & N & N & N & N & N & N & N \\
      \hline
      % 9 & - & - & - & - & - & - & - & - & - & - & - & - \\
      % \hline
      10 & N & N & N & N & N/A & N & N/A & \textcolor{red}{V} & N/A & N/A & N & N \\
      \hline
      11 & \textcolor{red}{V} & \textcolor{red}{V} & N & N & N/A & N & N/A & \textcolor{red}{V} & N/A & N/A & N & N \\
      \hline
      % 12 & - & - & - & - & - & - & - & - & - & - & - & - \\
      % \hline
      % 13 & - & - & - & - & - & - & - & - & - & - & - & - \\
      % \hline
      % 14 & - & - & - & - & - & - & - & - & - & - & - & - \\
      % \hline
      % 15 & - & - & - & - & - & - & - & - & - & - & - & - \\
      % \hline
      % 16 & - & - & - & - & - & - & - & - & - & - & - & - \\
      % \hline
      % 17 & - & - & - & - & - & - & - & - & - & - & - & - \\
      % \hline
      18 & \textcolor{red}{V} & N & N & N & N & N & \textcolor{red}{V} & \textcolor{red}{V} & \textcolor{red}{V} & \textcolor{red}{V} & \textcolor{red}{V}(ML) & \textcolor{red}{V}(ML) \\
      \hline
    \end{tabular}
    \end{adjustbox}
    \end{adjustwidth}
    
    \caption{DS1 to DS12: AARDROID, AMANDROID, and CRYPTOGUARD}
    \label{tab:ds1_to_ds12_chart}
\end{table*}

\textbf{\textit{Explanation of Table {\ref{tab:ds1_to_ds12_chart}}: }}
It is found by Aardroid that application no. 11 and 18 violated the property \textbf{DS1}-No Android Keystore usage for system credentials. However, only application no. 11 shows violation to the property \textbf{DS2}-Refrain from external storage. None of the applications show violation to \textbf{DS3}-Logging, \textbf{DS4}-Third party API leakage, \textbf{DS5}-UI/keyboard input cache, and \textbf{DS6}-Inter Process Communication (IPC) leakage. This means Aadroid found none of the application - which passed scanning - never sent sensitive data to device logs. Moreover, neither they pass sensitive data to third party APIs nor they leak data through IPC. They also do not leak data through keyboard cache. \textbf{DS7}-Sensitive data leakage through UI, \textbf{DS9}-Disable Screenshot UI, and \textbf{DS10}-Device Access Policy via pass-code setup- are all found to be violated by only application no. 18. However almost all of the application, except application no. 8, violated \textbf{DS8}-Disable auto-backup. Lastly, application no. 1, 5 and 18 violated the properties \textbf{DS11}-Avoiding Storage in Local Persistence, and \textbf{DS12}-Allowing Persistence through Data Encryption.\\
\vspace{12pt}

\begin{table*}[h]
    \begin{adjustwidth}{}{}
    \begin{adjustbox}{width=15cm}
    \begin{tabular}{|c|c|c|c|c|c|c|c|c|}
      \hline
        \textbf{App No.} & 
        \textbf{CRYPTO1} &
        \textbf{CRYPTO2} &
        \textbf{CRYPTO3} &
        \textbf{CRYPTO4} &
        \textbf{TLS1} &
        \textbf{TLS2} &
        \textbf{TLS3} &
        \textbf{TLS4} \\
      \hline
      1 & \textcolor{red}{V} & \textcolor{red}{V} & \textcolor{red}{V} & \textcolor{red}{V} & N & N & N & N \\
      \hline
      % 2 & - & - & - & - & - & - & - & - \\
      % \hline
      % 3 & - & - & - & - & - & - & - & - \\
      % \hline
      % 4 & - & - & - & - & - & - & - & - \\
      % \hline
      5 & N & N & \textcolor{red}{V} & \textcolor{red}{V} & N & N & N & N \\
      \hline
      % 6 & - & - & - & - & - & - & - & - \\
      % \hline
      % 7 & - & - & - & - & - & - & - & - \\
      % \hline
      8 & N & N & N & N & N & N & N & N \\
      \hline
      % 9 & - & - & - & - & - & - & - & - \\
      % \hline
      10 & N & N & \textcolor{red}{V} & \textcolor{red}{V} & N & \textcolor{red}{V} & \textcolor{red}{V} & N \\
      \hline
      11 & N & N & \textcolor{red}{V} & \textcolor{red}{V} & \textcolor{red}{V} & \textcolor{red}{V} & \textcolor{red}{V} & N \\
      \hline
      % 12 & - & - & - & - & - & - & - & - \\
      % \hline
      % 13 & - & - & - & - & - & - & - & - \\
      % \hline
      % 14 & - & - & - & - & - & - & - & - \\
      % \hline
      % 15 & - & - & - & - & - & - & - & - \\
      % \hline
      % 16 & - & - & - & - & - & - & - & - \\
      % \hline
      % 17 & - & - & - & - & - & - & - & - \\
      % \hline
      18 & N & \textcolor{red}{V} & \textcolor{red}{V} & \textcolor{red}{V} & N & N & N & \textcolor{red}{V} \\
      \hline
    \end{tabular}
    \end{adjustbox}
    \end{adjustwidth}
    
    \caption{CRYPTO1 to TLS4: AARDROID, AMANDROID, and CRYPTOGUARD}
    \label{tab:crypto1_to_tls4_chart}
\end{table*}

\textbf{\textit{Explanation of Table {\ref{tab:crypto1_to_tls4_chart}}: }}
The Descriptions of the properties on Table \textcolor{red}{3} are as follows : \textbf{CRYPTO1}-Hard-coded Cryptographic keys, \textbf{CRYPTO2}-Appropriate Cryptographic Configuration (e..g. insufficient key lengths, usage of insecure modes (such as ECB Mode), inclusion of predictable or static Initialization Vectors (IVs) in symmetric ciphers, and an insufficient number of iterations in Password-Based Encryption (PBE)), \textbf{CRYPTO3}-Usage of non-deprecated cryptographic algorithm, \textbf{CRYPTO4}-Correct selection and setup of random number generators (insecure PRNG), \textbf{TLS1}-SDK / application employs TLS network connections (HTTP), \textbf{TLS2}-No Broken TLS configuration,  \textbf{TLS3}-Appropriate validation of endpoint certificates, and  \textbf{TLS4}-No certificate pinning. Application no. 1 violates all of \textbf{CRYPTO1-CRYPTO4} properties while Application no. 5 and 10 both violate only \textbf{CRYPTO3} \& \textbf{CRYPTO4}. In addition to that, app no. 10 also show violation to \textbf{TLS2} \& \textbf{TLS3}. Moreover, Application no. 11 violates the properties \textbf{CRYPTO1-CRYPTO2} \& \textbf{TLS1-TLS3}. Application no. 18 is the only one that show violation to \textbf{TLS4}. Beside that it also violates the properties \textbf{CRYPTO2-CRYPTO4}. However, application no. 8 does not show any violation to any of the Table {\ref{tab:crypto1_to_tls4_chart}} properties.\\
\vspace{12pt}

\begin{table*}[h]
    \centering
    % \small
    \begin{adjustwidth}{}{}
    \begin{adjustbox}{width=15cm}
    \begin{tabular}{|c|c|c|c|c|c|c|c|c|}
      \hline
        \textbf{App No.} & 
        \textbf{PLAT1} &
        \textbf{PLAT2} &
        \textbf{PLAT3} &
        \textbf{PLAT4} &
        \textbf{PLAT5} &
        \textbf{PLAT6} & 
        \textbf{PLAT8}&
        \textbf{PLAT7} \\
      \hline
      1 & N & N & N & N/A & N/A & N/A & N/A & N/A\\
      \hline
      2 & - & \textcolor{red}{V} & \textcolor{red}{V} & \textcolor{red}{V} & \textcolor{red}{V} & N & \textcolor{red}{V} & \textcolor{red}{V} \\
      \hline
      % 3 & - & - & - & - & - & - & - & - \\
      % \hline
      % 4 & - & - & - & - & - & - & - & -\\
      % \hline
      5 & N & N & N & N/A & N/A & N/A & N/A & N/A \\
      \hline
      % 6 & - & - & - & - & - & - & - & -\\
      % \hline
      % 7 & - & - & - & - & - & - & - & -\\
      % \hline
      8 & N & \textcolor{red}{V} & \textcolor{red}{V} & \textcolor{red}{V} & \textcolor{red}{V} & N & N & N\\
      \hline
      % 9 & - & - & - & - & - & - & - & -\\
      % \hline
      10 & N & N & N & N/A & N/A & N/A & N/A & N/A\\
      \hline
      11 & N & N & N & N & N/A & N & N/A & N/A \\
      \hline
      12 & - & N & N & \textcolor{red}{V} & \textcolor{red}{V} & N & N & N \\
      \hline
      % 13 & - & - & - & - & - & - & - & - \\
      % \hline
      % 14 & - & - & - & - & - & - & - & -\\
      % \hline
      % 15 & - & - & - & - & - & - & - & -\\
      % \hline
      16 & - & N & N & \textcolor{red}{V} & \textcolor{red}{V} & N & N & N \\
      \hline
      17 & - & N & N & \textcolor{red}{V} & \textcolor{red}{V} & \textcolor{red}{V} & N & N \\
      \hline
      18 & N & N & N & \textcolor{red}{V} & N & N & \textcolor{red}{V} & \textcolor{red}{V} \\
      \hline
    \end{tabular}
    \end{adjustbox}
    \end{adjustwidth}
    
    \caption{PLAT1 to PLAT7: AARDROID, AMANDROID, and CRYPTOGUARD \\ 
    }\textbf{Note : PLAT8 is placed before PLAT7 because this is how the output was generated}
    \label{tab:plat1_to_plat7_chart}
\end{table*}

\textbf{\textit{Explanation of Table {\ref{tab:plat1_to_plat7_chart}}: }}
Following are Table \textcolor{red}{{\ref{tab:plat1_to_plat7_chart}}} field descriptions: (\textbf{PLAT1}-SDK / application must refrain from dangerous permission, \textbf{PLAT2}-Avoid sensitive information exposure via custom URL schemes, \textbf{PLAT3}-Avoid exposure of sensitive functionalities through IPC, \textbf{PLAT4}-Security-conscious configuration should complement WebView , including the deactivation of JavaScript, \textbf{PLAT5}-Refrain from unsafe protocol, \textbf{PLAT6}-Refrain from exposing a JavaScript bridge, \textbf{PLAT7}-Shield against screen overlay attacks, and \textbf{PLAT8}-Learn the web cache prior to closure. Among the application which the Aardroid is able to scan, none of them show any violation to \textbf{PLAT1} property. However,  application no. 2 \& 8 both violates \textbf{PLAT2-PLAT5}, meaning sensitive information leakage through URL and IPC and use of unsafe protocol with activation of JavaScript in WebView which might be prone to XSS. Additionally, application no. 2 shows violation to \textbf{PLAT7} \& \textbf{PLAT8} along with application no. 18, indicating the possibility of screen overlay and web cache attacks. Moreover, application no. 12, 16 \& 17 violate the properties \textbf{PLAT4} \& \textbf{PLAT5}, with application no. 18 violating \textbf{PLAT4}. Lastly, the only application to violate \textbf{PLAT6} is no. 18, indicating the possibility of JavaScript bridge exposure.\\

\begin{figure}[htbp]
\centering
\includegraphics[width=0.5\textwidth]{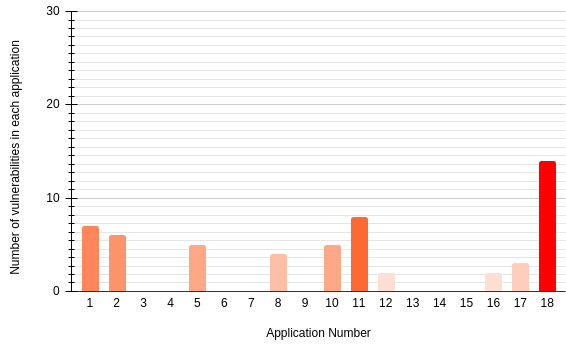}
\caption{Bar chart based on vulnerabilities found on each applications}
\label{fig:Bar chart based on vulnerabilities found on each applications}
\end{figure}

The bar chart in figure \textcolor{red}{\ref{fig:Bar chart based on vulnerabilities found on each applications}} gives a birds eye view of the applications which are vulnerable to different vulnerabilities as found in the tables \textcolor{red}{\ref{tab:ds1_to_ds12_chart}}, \textcolor{red}{\ref{tab:crypto1_to_tls4_chart}} and \textcolor{red}{\ref{tab:plat1_to_plat7_chart}} which are tabulated from the output of scanned result of Aardroid. Hence, it is  easy to visualize that how many security threats does each application impose and how severe is the condition with respect to each applications.\\

\begin{figure}[htbp]
\centering
\includegraphics[width=0.5\textwidth]{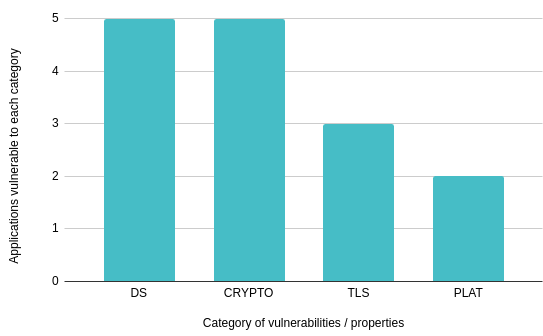}
\caption{Bar chart based on application vulnerable to each category}
\label{fig:Bar chart based on application vulnerable to each category}
\end{figure}

By observing the bar chart it is very clear that the application no. 18, which is the payment SDK by SSL Commerz, violates almost half of the properties we considered in our investigation while the next most violating applications after no.18 are application no. 1, 2 and 11 which are basically the Mobile Banking and Wallet Application. Moreover, a bar chart is also plotted to visualize the number of applications containing vulnerabilities in each vulnerability group, as shown in fig \ref{fig:Bar chart based on application vulnerable to each category}, to understand the distribution of the statistics.

\section{Results obtained through manual examination: }

% According to the flowchart in Fig. \textcolor{red}{\ref{fig:process_flow_chart}}, 

After scanning the application and observing the result we are supposed to manually verify whether the mentioned violation states are True positive, False positive, True negative or False negative. On doing source code analysis we discovered that some of the application do not perform bidirectional authentication; they specifically ignored the server side client certificate verification. This could be a potential vulnerability that could be exploited if any custom request is sent with remote code injection.\\

One of the many surprising findings to mention is that some of the banking application make use of the One Time Pin (OTP) for verification. Some of these application uses the Google API service \textbf{Automatic SMS Verification with the SMS Retriever API}. So bypassing it is a bit tricky. However, with close inspection we can find that the SMS actually follows certain specific pattern which triggers the API service. Once we bypass the SMS trick, we are able to register from the emulator which causes an OTP to be sent to the mobile device. If we somehow are able to forward the SMS or copy it to the emulator's system settings and send it from the dummy SMS generator to the emulator, the service gets triggered, hence the OTP gets retrieved automatically bypassing the security measure in a legal way. This indicates there is the absence of device binding which should have restricted the execution of the application outside of specific devices. \\

Further into our investigation, we tried to exploit the other part of the potential vulnerabilities through source code modification. However, some of the vendors pushed frequent updates to  their applications which made our previous findings and workings irrelevant on the applications. The challenges we faced is discussed in the next section in detail. While some of the applications were based on pure APK files, others which were not decompilable using Aardroid was due to they being XAPK format which is the new format for android application, containing list of APKs inside. But we tried to decompile the files using Apktool on which we were able to do successfully. Besides Apktool, any sort of zip extraction tool works fine since APK file is nothing but a zipped format of the android application containging the meta data, static resources and the byte code format of the source code. However, the inside of a decompiled XAPK file is not the same as a decompiled APK file. The source codes are not there on the same paths anymore. Hence, our work did not work on those applications using XAPK format.\\

\section{Challenges}
\label{challenges}
Throughout our investigation, we encountered a series of significant challenges that underscore the complexity of the task at hand. One of the primary obstacles was the scarcity of open-source scanners tailored to our specific scope. The absence of readily available tools designed to delve into the intricacies of APK analysis impede our progress, necessitating the exploration of alternative scanning approaches. Additionally, the dearth of well-structured open-source documentation, apart from the official resources, posed a significant challenge. The limited availability of comprehensive materials hindered our efforts to thoroughly comprehend and scrutinize the intricacies of the targeted APKs. Compounding these challenges, we identified a notable shortage of researchers actively engaged in this specialized sector, particularly within our region. This scarcity of expertise and collaboration further compounded the intricacies of our investigation. As we navigate these hurdles, we are compelled to forge new paths, draw insights from available resources, and innovate novel approaches to overcome these challenges and drive our investigation forward. \\

One of the major issues we faced in our research work was code Obfuscation. Most of the tools did not work with the obfuscated codes. Majority of the research works were done based on codes not having obfuscation. Till date Obfuscation is a good technique which works well. But it might be true that the application using obfuscation to hide the code structure might not be following industry standards or have an unnoticed region of potential vulnerabilities. Hence attackers with good knowledge might be able to exploit the system without much struggle. Hence, we did not want to leave the obfuscated application. We rather tried to analyze it manually which consumed huge portions of the time since we fall into frequent rabbit holes.\\

But the main challenge throughout the investigation was shortage of time. It could be rather said in this way that we had time but the majority of the time went in learning new stuff, in searching and studying existing work, but most importantly it took more time to understand the working principles of certain concepts and trying to find methods to analyze them. For example, the approach we used in Section \ref{sec:investigation} to do source code manipulation did not work anymore on the second update of one of the applications we have been working with. By the time we were finding other approaches to defeat the new system patch the vendors changed the APK format to XAPK format as mentioned in the previous section. The problem was not the format, rather there are no available resources regarding the XAPK format. We searched in forums, research papers, blogs and development sites; but we could not find any bits of particle for this one.  \\

By the time we reach a certain destination, the concept used so far becomes obsolete. This is due to the lack of research in this sector that causes the rise in coming up with a new method. Though it might be good for the development side and profitable for the businesses, if the security researcher in this sector does not try to match up with the pace of development, then future disaster like that happened in Bangladesh Bank Heist is not very far away.\\

Since it takes years of experience to make a creative approach to exploit existing vulnerability to get a zero day, we urge researchers to come forward and invest their time and interest more into this sector.

\section{Conclusions}
\label{conclusions}

In this paper we tried to investigate the current android applications that are being used in the daily chores of the Bangladeshi locals. We tried to analyze the applications for possible potential security vulnerabilities and tried to penetrate through them to compromise the security system for the betterment of the user. Despite of the fact that through automated scanning we were able to find the presence of potential vulnerabilities in the applications, but due to the vendor specific developers following the android ecosystem and developers manual, everytime Google makes new update to the libraries, many of the developers follow the strategy of updating their code base, minimizing the risks to be encountered with potential vulnerability and reducing the chance for any malicious user (attacker) trying to compromise the system. Hence, even though they might not implement any separate special techniques to secure their system, following the guidelines also ensures their safety in the best possible way. Most importantly, we tried to provide a documented guideline and framework on how anyone, whether they being a part of any organization or an individual, can do security analysis on the android applications for both learning purposes and safety purposes for the business. As more researchers become interested in this sector, the system will eventually become much more secure.\\

\textit{\textbf{Recommendation:}\\
Application vendor can make their public key of their certificate publicly available to some public database so that it could be downloaded to the client devices in order to cross check with the downloaded APK to verify whether the application has been actually signed by the correct and trusted authority or not.}

\section{Future Work}
Since not much work is done on the obfuscation techniques of the android application source code, there is high potential to look into this area of android security. Moreover, XAPK format is also a well known new term which does not have any research work on, at least that is how far we have found through our research. So working in this region is also logical to get great output. Besides these, we are also planning to work on the POC of the potential vulnerabilities of our research work. In addition to that, working towards the development of an open source scanner is also in our concern along with a parser that will work on the byte code directly to look for method and class signatures so that code manipulation can become easier.\\

\textit{\textbf{Availability : }
The source code for the customized version of AARDroid is publicly available at  \href{https://github.com/shahriar-hasan-mickey/aardroid\_modified}{github.com/shahriar-hasan-mickey/aardroid\_modified}.}

\bibliographystyle{plain} % We choose the "plain" reference style
\bibliography{bibliography/the_main} % Entries are in the refs.bib file

\begin{thebibliography}{10}

\bibitem{mobsf}
Ajin Abraham.
\newblock {Mobile Security Framework (MobSF)}.
\newblock Retrieved \today{} \url{https://github.com/MobSF/Mobile-Security-Framework-MobSF}, 2023.

\bibitem{adascalitei}
Ioan ADĂSCĂLIȚEI.
\newblock Smartphones and iot security.
\newblock {\em Informatica Economică}, 23(2):0, 2019.

\bibitem{impactstudy}
Md. Akhtaruzzaman, Md.~Ezazul Islam, Md.~Syedul Islam, Shibli Rubayat-Ul Islam, Mohammad Rakib~Uddin Bhuiyan, Mohammad Tareq, and Md. Jahir~Uddin Palas.
\newblock An impact study on mobile financial services (mfss) in bangladesh.
\newblock \url{https://www.bb.org.bd/pub/special/impact_mfs_27092018.pdf}, 2017.

\bibitem{Akter2021}
Sharmin Akter, Najma Kabir, and Tahmina Reza.
\newblock Unfolding factors behind internet-banking adoption in bangladesh: An extension of utaut2 with perceived security and trust.
\newblock {\em The Journal of Management Theory and Practice}, 2(2), August 2021.

\bibitem{alzadjali}
Buthaina~Mohammed AL-Zadjali.
\newblock A critical evaluation of vulnerabilities in android os: (forensic approach).
\newblock {\em International Journal of Computer Applications}, 130(5), November 2015.

\bibitem{radare2}
Sergi Alvarez.
\newblock {Radare2: Libre Reversing Framework for Unix Geeks}.
\newblock Retrieved \today{} \url{https://github.com/radareorg/radare2}, 2023.

\bibitem{amarante}
João Amarante and João~Paulo Barros.
\newblock Exploring usb connection vulnerabilities on android devices - breaches using the android debug bridge.
\newblock In {\em Proceedings of the 14th International Joint Conference on e-Business and Telecommunications (ICETE 2017) - Volume 4: SECRYPT}, pages 572--577, 2017.

\bibitem{flowdroid}
Steven Arzt, Siegfried Rasthofer, Christian Fritz, Eric Bodden, Alexandre Bartel, Jacques Klein, Yves Le~Traon, Damien Octeau, and Patrick McDaniel.
\newblock Flowdroid.
\newblock {\em ACM SIGPLAN Notices}, 49:259--269, 06 2014.

\bibitem{reverseengineering}
Syeda~Warda Asher, Sadeeq Jan, George Tsaramirsis, Fazal~Qudus Khan, Abdullah Khalil, and Muhammad Obaidullah.
\newblock Reverse engineering of mobile banking applications.
\newblock {\em Computer Systems Science \& Engineering}, 2021.

\bibitem{bartel}
Alexandre Bartel, Jacques Klein, Martin Monperrus, and Yves Le~Traon.
\newblock Automatically securing permission-based software by reducing the attacksurface: An application to android.
\newblock {\em IEEE/ACM International Conference on Automated Software Engineering}, pages 274--277, 09 2012.

\bibitem{burpsuit}
Burpsuit.
\newblock {Burp Suite : Application Security Testing Software - PortSwigger}.
\newblock Retrieved \today{} \url{https://portswigger.net/burp}, 2024.

\bibitem{tapjacking}
Vanessa Cooper.
\newblock {\em Tapjacking Threats and Mitigation Techniques for Android Applications}.
\newblock PhD thesis, 05 2014.

\bibitem{side_channel}
Patrick Cronin, Xing Gao, Chengmo Yang, and Haining Wang.
\newblock {Charger-Surfing}: Exploiting a power line {Side-Channel} for smartphone information leakage.
\newblock In {\em 30th USENIX Security Symposium (USENIX Security 21)}, pages 681--698. USENIX Association, August 2021.

\bibitem{droidcap}
Abdallah Dawoud and Sven Bugiel.
\newblock Droidcap: Os support for capability-based permissions in android.
\newblock 01 2019.

\bibitem{androguard}
Anthony Desnos, Geoffroy Gueguen, and Paul~Sabatier University.
\newblock {AndroGuard}.
\newblock Retrieved \today{} \url{https://github.com/androguard/androguard}, 2023.

\bibitem{reaper}
Michalis Diamantaris, Elias~P. Papadopoulos, Evangelos~P. Markatos, Sotiris Ioannidis, and Jason Polakis.
\newblock Reaper: Real-time app analysis for augmenting the android permission system.
\newblock In {\em Proceedings of the Ninth ACM Conference on Data and Application Security and Privacy}, CODASPY '19, page 37–48, New York, NY, USA, 2019. Association for Computing Machinery.

\bibitem{xposed}
Drupal.
\newblock Xposed module repository.
\newblock Retrieved \today{} from \url{https://repo.xposed.info/}, 2024.

\bibitem{jadx}
Emmanuel Dupuy.
\newblock {JADX}.
\newblock Retrieved \today{} \url{https://github.com/skylot/jadx}, 2023.

\bibitem{taintdroid}
William Enck, Peter Gilbert, Seungyeop Han, Vasant Tendulkar, Byung-Gon Chun, Landon~P. Cox, Jaeyeon Jung, Patrick McDaniel, and Anmol~N. Sheth.
\newblock Taintdroid: An information-flow tracking system for realtime privacy monitoring on smartphones.
\newblock {\em ACM Trans. Comput. Syst.}, 32(2), jun 2014.

\bibitem{amandroid}
Xinming~Ou Fengguo~Wei, Sankardas~Roy and Robby.
\newblock {\em Amandroid: A Precise and General Inter-component Data Flow Analysis Framework for Security Vetting of Android Apps}.
\newblock PhD thesis, University of South Florida, 2014.

\bibitem{fortify}
Fortify.
\newblock {Fortify Static Code Analyzer}.
\newblock Retrieved \today{} \url{https://www.microfocus.com/documentation/fortify-static-code-analyzer-and-tools/2310/}, 2023.

\bibitem{fratantonio17:cloakdagger}
Yanick Fratantonio, Chenxiong Qian, Simon Chung, and Wenke Lee.
\newblock {Cloak and Dagger: From Two Permissions to Complete Control of the UI Feedback Loop}.
\newblock In {\em Proceedings of the IEEE Symposium on Security and Privacy (Oakland)}, San Jose, CA, May 2017.

\bibitem{frida}
Frida.
\newblock {Frida : A Reversing Engineering Tool}.
\newblock Retrieved \today{} \url{https://frida.re/}, 2024.

\bibitem{gdroid}
Han Gao, Shaoyin Cheng, and Weiming Zhang.
\newblock Gdroid: Android malware detection and classification with graph convolutional network.
\newblock {\em Comput. Secur.}, 106(C), jul 2021.

\bibitem{ghidra}
Ghidra.
\newblock {Ghidra : A Reversing Engineering Framework}.
\newblock Retrieved \today{} \url{https://github.com/NationalSecurityAgency/ghidra}, 2024.

\bibitem{google_android_1}
Google.
\newblock {Opcodes}.
\newblock Retrieved \today{} from \url{https://developer.android.com/reference/dalvik/bytecode/Opcodes.html}, 2019.

\bibitem{google_android_2}
Google.
\newblock Dalvik bytecode.
\newblock Retrieved \today{} from \url{https://source.android.com/docs/core/runtime/dalvik-bytecode}, 2022.

\bibitem{androidstudio}
{Google} and {JetBrains}.
\newblock {Android Studio}.
\newblock Retrieved \today{} from \url{https://developer.android.com/studio}, 2023.

\bibitem{droidsafe}
Michael Gordon, Kim deokhwan, Jeff Perkins, Limei Gilham, Nguyen Nguyen, and Martin Rinard.
\newblock Information-flow analysis of android applications in droidsafe.
\newblock 01 2015.

\bibitem{iccta}
Li~Li, Alexandre Bartel, Tegawendé~F. Bissyandé, Jacques Klein, Yves Le~Traon, Steven Arzt, Siegfried Rasthofer, Eric Bodden, Damien Octeau, and Patrick McDaniel.
\newblock Iccta: Detecting inter-component privacy leaks in android apps.
\newblock In {\em 2015 IEEE/ACM 37th IEEE International Conference on Software Engineering}, volume~1, pages 280--291, 2015.

\bibitem{androbugs}
Yu-Cheng Lin.
\newblock {AndroBugs}.
\newblock Retrieved \today{} \url{https://github.com/AndroBugs/AndroBugs_Framework}, 2023.

\bibitem{qark}
LinkedIn.
\newblock {QARK (Quick Android Review Kit)}.
\newblock Retrieved \today{} \url{https://github.com/linkedin/qark}, 2019.

\bibitem{aardroid-acsac22}
Samin~Yaseer Mahmud, K.~Virgil English, Seaver Thorne, William Enck, Adam Oest, and Muhammad Saad.
\newblock {Analysis of Payment Service Provider SDKs in Android}.
\newblock In {\em {Proceedings of the Annual Computer Security Applications Conference (ACSAC)}}, 2022.

\bibitem{bBank}
Mobile Financial~Services (MFS).
\newblock {Mobile Financial Services (MFS) comparative summary statement of May, 2023 and June, 2023}.
\newblock Retrieved \today{} \url{https://www.bb.org.bd/en/index.php/financialactivity/mfsdata}, 2023.

\bibitem{masvs}
Bernhard Mueller, Sven Schleier, Jeroen Willemsen, and Carlos Holguera.
\newblock {MASVS : Mobile Application Security Verification Standard}.
\newblock Retrieved \today{} from \url{https://owasp.org/www-project-mobile-top-10/2016-risks/}, 2023.

\bibitem{rootAvd}
newbit.
\newblock {rootAVD}.
\newblock Retrieved \today{} \url{https://github.com/newbit1/rootAVD}, 2023.

\bibitem{epicc}
Damien Octeau, Patrick McDaniel, Somesh Jha, Alexandre Bartel, Eric Bodden, Jacques Klein, and Yves Le~Traon.
\newblock Effective inter-component communication mapping in android with epicc: an essential step towards holistic security analysis.
\newblock In {\em Proceedings of the 22nd USENIX Conference on Security}, SEC'13, page 543–558, USA, 2013. USENIX Association.

\bibitem{gsma}
Kenechi Okeleke.
\newblock Achieving mobile-enabled digital inclusion in bangladesh.
\newblock \url{https://tinyurl.com/33pnzdx6}, 2021.

\bibitem{parvez2021}
Nasim Parvez, Tamjidul~Haque Chowdhury, Shahina~Sultana Urmi, and Kazi~Abu Taher.
\newblock Prospects of internet of things for bangladesh.
\newblock In {\em Proceedings of the IEEE International Conference on ICT for Sustainable Development (ICICT4SD)}. IEEE, 2021.

\bibitem{dypoldroid}
Carlos~E. Rubio-Medrano, Matthew Hill, Luis~M. Claramunt, Jaejong Baek, and Gail-Joon Ahn.
\newblock Dypoldroid: Protecting users and organizations from permission-abuse attacks in android.
\newblock 2021.

\bibitem{internetBanking}
Bipasha Sarker, Bidisha Sarker, Prajoy Podder, and Md~Alam~Robel.
\newblock Progression of internet banking system in bangladesh and its challenges.
\newblock International Journal of Computer Applications (0975 – 8887) Volume 177 – No. 29, January 2020, 2020.

\bibitem{bBHeist}
Mathew~J. Schwartz.
\newblock Bangladesh bank attackers hacked swift software, 2016.

\bibitem{sonarqube}
SonarQube.
\newblock {SonarQube}.
\newblock Retrieved \today{} \url{https://www.sonarqube.org/}, 2023.

\bibitem{top10}
Milan~Singh Thakur, Alaeddine MESBAHI, Kunwar Atul, and Mohamed Benchikh.
\newblock {Top 10 Mobile Risks}.
\newblock Retrieved \today{} \url{https://owasp.org/www-project-mobile-top-10/}, 2023.

\bibitem{apktool}
Connor Tumbleson.
\newblock {Apktool}.
\newblock Retrieved \today{} \url{https://github.com/iBotPeaches/Apktool}, 2023.

\bibitem{pcissc}
Branden~R. Williams, David Mundhenk, Yves~B. Desharnais, Anton Chuvakin, Arthur~B. Cooper~Jr., Nicki Carter, Jim Seaman, Jeff Hall, and Francois Desharnais.
\newblock {Payment Card Industry's Data Security Standard (PCI-DSS)}.
\newblock \url{https://docs-prv.pcisecuritystandards.org/PCI\%20DSS/Standard/PCI-DSS-v4_0.pdf}, 2022.

\bibitem{vetdroid}
Yuan Zhang, Min Yang, Bingquan Xu, Zhemin Yang, Guofei Gu, Peng Ning, X.~Sean Wang, and Binyu Zang.
\newblock Vetting undesirable behaviors in android apps with permission use analysis.
\newblock page 611–622, New York, NY, USA, 2013. Association for Computing Machinery.

\bibitem{BorderPatrol}
O.~Zungur, G.~Suarez-Tangil, G.~Stringhini, and M.~Egele.
\newblock Borderpatrol: Securing byod using fine-grained contextual information.
\newblock In {\em 2019 49th Annual IEEE/IFIP International Conference on Dependable Systems and Networks (DSN)}, pages 460--472, Los Alamitos, CA, USA, jun 2019. IEEE Computer Society.

\end{thebibliography}

% \begin{table}[h]
%     \centering
%     \begin{tabular}{|c|l|c|}
%     \hline
%     App No. & Permission required \\
%     \hline
%     \multirow{7}{*}{1 &    |  |   \\
%     % \hline
%                         &   |  |    \\
%                         &  |   \\
%                         &  |  |  |   \\
%                         &  |   \\
%                         &  |  |   \\
%                         &  |     \\
%                         & \\

\begin{table*}
    \centering
    % % \small
    \begin{adjustwidth}{-2cm}{}
    \begin{adjustbox}{width=19cm}
    \begin{tabular}{|c|c|c|c|c|c|c|c|c|c|c|c|c|c|c|c|c|c|}
      \hline
      
        \textbf{Permissions}& 1 & 2 & 3 & 4 & 5 & 6 & 7 & 8 & 9 & 10 & 11 & 12 & 13 & 14 & 15 & 16 & 17\\
        
      \hline
      
      ACCESS\_NETWORK\_STATE & \cmark & \cmark & \cmark  & \cmark & \cmark & \cmark & \cmark & \cmark & \cmark & \cmark & \cmark & \cmark & \cmark & \cmark & \cmark & \cmark & \cmark \\
      
      \hline

      SYSTEM\_ALERT\_WINDOW &  &  &  &  &  & \cmark &  &  &  &  & \ &  &  &  &  &  & \\
      
      \hline
      
      SCHEDULE\_EXACT\_ALARM & \cmark &  &  &  &  &  &  &  &  &  &  &  &  &  & \cmark &  & \\
      
      \hline
      
      VIBRATE& \cmark &  &  &  & \cmark &  &  & \cmark &  & \cmark &  & \cmark &  & \cmark & \cmark &  & \\
      
      \hline
      
      USE\_FULL\_SCREEN\_INTENT & \cmark &  &  &  &  &  &  &  &  &  &  & \cmark &  &  &  &  &\\
      
      \hline
      
      INTERNET& \cmark & \cmark & \cmark & \cmark & \cmark & \cmark & \cmark & \cmark & \cmark & \cmark & \cmark & \cmark & \cmark & \cmark & \cmark & \cmark & \cmark \\
      
      \hline
      
      GET\_TASKS& \cmark &  &  & \cmark &  &  &  &  &  &  &  &  &  &  &  &  &\\
      
      \hline
      
      READ\_MEDIA\_IMAGES& \cmark &  &  &  &  &  &  &  & \cmark &  &  &  & \cmark &  &  &  &\\
      
      \hline
      
      READ\_EXTERNAL\_STORAGE & \cmark & \cmark & \cmark & \cmark &  & \cmark & \cmark & \cmark & \cmark & \cmark & \cmark & \cmark & \cmark & \cmark & \cmark & \cmark & \cmark \\
      
      \hline

      CALL\_PHONE &  &  & &  &  & \cmark &  & \cmark &  & \cmark &  &  &  &  &  & \cmark &\\
      
      \hline
      
      READ\_CONTACTS& \cmark &  & \cmark  &  & \cmark &  & \cmark &  & \cmark &  & \cmark &  & \cmark & \cmark &  &  &\\
      
      \hline
      
      RECORD\_AUDIO& \cmark &  &  &  &  &  &  &  & \cmark &  & \cmark &  &  &  &  &  &\\
      
      \hline
      
      WAKE\_LOCK& \cmark & \cmark & \cmark & \cmark & \cmark & \cmark & \cmark & \cmark & \cmark & \cmark & \cmark & \cmark & \cmark & \cmark & \cmark & \cmark & \cmark\\
      
      \hline
      
      REORDER\_TASKS& \cmark &  &  &  &  &  &  &  &  &  &  & &  & \cmark &  &  &\\
      
      \hline
      
      ACCESS\_FINE\_LOCATION& \cmark &  & \cmark & \cmark & \cmark &  & \cmark & \cmark & \cmark &  & \cmark &  & \cmark & \cmark & \cmark &  &\\
      
      \hline
      
      RECEIVE\_BOOT\_COMPLETED & \cmark & \cmark &  & \cmark &  &  & \cmark &  &  &  & \cmark & \cmark & \cmark & \cmark & \cmark &  & \cmark\\
      
      \hline
      
      USE\_FINGERPRINT & \cmark &  &  & \cmark &  &  & \cmark & \cmark & \cmark &  &  &  &  &  &  &  &\\
      
      \hline
      
      READ\_PHONE\_STATE& \cmark &  & \cmark &  & \cmark &  & \cmark & \cmark &  &  & \cmark &  &  & \cmark &  &  &\\
      
      \hline
      
      FOREGROUND\_SERVICE& \cmark & \cmark & & \cmark &  & \cmark & \cmark &  &  &  & \cmark &  & \cmark & \cmark & \cmark &  & \cmark\\
      
      \hline
      
      ACCESS\_WIFI\_STATE& \cmark & \cmark & \cmark & \cmark &  & \cmark & \cmark & \cmark &  &  & \cmark &  & \cmark & \cmark & \cmark & \cmark & \cmark\\
      
      \hline

        WRITE\_EXTERNAL\_STORAGE& \cmark & \cmark & \cmark & \cmark & \cmark & \cmark & \cmark & \cmark & \cmark & \cmark & \cmark & \cmark & \cmark & \cmark & \cmark & \cmark & \cmark \\
        
        \hline

        USE\_BIOMETRIC& \cmark &  &  & \cmark &  &  & \cmark &  & \cmark &  &  &  &  &  &  &  &\\

        \hline
        
        ACCESS\_COARSE\_LOCATION& \cmark &  &  & \cmark & \cmark & \cmark &  & \cmark & \cmark & \cmark &  &  &  & \cmark & \cmark & \cmark &\\

        \hline
        
        POST\_NOTIFICATIONS& \cmark &  &  &  & \cmark &  &  &  & \cmark & \cmark &  &  &  & \cmark & \cmark &  &\\

        \hline
        
        CAMERA& \cmark & \cmark & \cmark & \cmark & \cmark & \cmark & \cmark & \cmark & \cmark & \cmark & \cmark & \cmark & \cmark & \cmark & \cmark & \cmark & \cmark\\

        \hline

        GET\_ACCOUNTS &  & \cmark &  & \cmark &  &  & \cmark &  &  & \cmark & \cmark &  & \cmark &  &  &  & \cmark\\

        \hline

        INSTALL\_PACKAGES &  & \cmark &  & \cmark &  &  & \cmark &  &  &  & \cmark &  & \cmark &  &  &  & \cmark\\

        \hline

        MANAGE\_ACCOUNTS &  & \cmark &  & \cmark &  &  & \cmark &  &  &  & \cmark &  & \cmark &  &  &  & \cmark\\

        \hline

        CHANGE\_WIFI\_MULTICAST\_STATE &  & \cmark &  & \cmark &  &  & \cmark &  &  &  & \cmark &  & \cmark &  &  &  & \cmark\\

        \hline

        USE\_CREDENTIALS &  & \cmark &  & \cmark &  &  & \cmark &  &  &  & \cmark &  & \cmark &  &  &  & \cmark\\

        \hline

        WRITE\_SYNC\_SETTINGS &  & \cmark & \cmark & \cmark &  &  & \cmark &  &  &  & \cmark &  & \cmark &  &  &  & \cmark\\

        \hline

        REQUEST\_INSTALL\_PACKAGES &  & \cmark & \cmark & \cmark &  &  & \cmark &  &  &  & \cmark &  & \cmark &  &  &  & \cmark\\

        \hline

        REQUEST\_DELETE\_PACKAGES &  & \cmark &  & \cmark &  &  & \cmark &  &  &  & \cmark &  & \cmark &  &  &  & \cmark\\

        \hline

        READ\_SYNC\_STATS &  & \cmark & & \cmark &  &  & \cmark &  &  &  & \cmark &  & \cmark &  &  &  & \cmark\\

        \hline

        AUTHENTICATE\_ACCOUNTS &  & \cmark &  & \cmark &  &  & \cmark &  &  &  & \cmark &  & \cmark &  &  &  & \cmark\\

        \hline

        READ\_SYNC\_SETTINGS &  & \cmark &  & \cmark &  &  & \cmark &  &  &  & \cmark &  & \cmark &  &  &  & \cmark\\

        \hline

        READ\_PHONE\_NUMBERS &  &  & \cmark &  &  &  &  &  &  &  &  &  &  &  &  &  &\\
        
        \hline 

        WRITE\_MEDIA\_IMAGES &  &  & \cmark &  &  &  &  &  &  &  &  &  &  &  &  &  &\\

        \hline

        ACCESS\_ADSERVICES\_ATTRIBUTION &  &  &  &  & \cmark & \cmark &  &  &  &  &  &  &  &  &  &  &\\

        \hline

         ACCESS\_ADSERVICES\_AD\_ID &  &  &  &  & \cmark & \cmark &  &  &  &  &  &  &  &  &  &  &\\

        \hline

        ACCESS\_ADSERVICES\_TOPICS &  &  &  &  &  & \cmark &  &  &  &  &  &  &  &  &  &  &\\

        \hline

        CONNECTIVITY\_ACTION &  &  &  &  &  &  &  & \cmark &  &  &  &  &  &  &  &  &\\

        \hline

        WRITE\_CONTACTS &  &  &  &  &  &  &  & \cmark &  & \cmark &  &  &  &  & \cmark &  &\\

        \hline

        BLUETOOTH &  &  &  &  &  &  &  &  & \cmark & \cmark &  &  &  &  &  &  &\\

        \hline

        BLUETOOTH\_CONNECT &  &  &  &  &  &  &  &  &  & \cmark &  &  &  &  &  &  &\\

        \hline

        READ\_MEDIA\_VIDEO &  &  &  &  &  &  &  &  & \cmark &  &  &  &  &  &  &  &\\

        \hline

        WRITE\_SETTINGS &  &  &  &  &  &  &  &  &  & \cmark &  &  &  &  &  &  &\\

        \hline

        READ\_MEDIA\_AUDIO &  &  &  &  &  &  &  &  &  & \cmark &  &  &  &  &  &  &\\

        \hline

        NFC &  &  &  &  &  &  &  &  &  & \cmark &  &  &  &  &  &  &\\

        \hline

        READ\_APP\_BADGE &  &  &  &  &  &  &  &  &  &  &  & \cmark &  &  &  &  &\\

        \hline

        RECEIVE\_SMS &  &  &  &  &  &  &  &  &  &  &  &  &  & \cmark &  &  &\\

        \hline

        READ\_SMS &  &  &  &  &  &  &  &  &  &  &  &  &  & \cmark &  &  &\\

        \hline

        REQUEST\_IGNORE\_BATTERY\_OPTIMIZATIONS &  &  &  &  &  &  &  &  &  &  &  &  &  &  & \cmark &  &\\

        \hline

        ACCESS\_COARSE\_UPDATES &  &  &  &  &  &  &  &  &  &  &  &  &  &  & \cmark &  &\\

        \hline

    \end{tabular}
    \end{adjustbox}
    \end{adjustwidth}
    
    \caption{Required permission by target applications}
    \label{tab:Required permission by target applications}
\end{table*}

\end{document}